\documentclass{PoS}
\usepackage{graphicx}

\newcommand{\citet}{\cite}
\newcommand{\citep}{\cite}

\title{Pulsar glitch substructure and pulsar interiors}

\ShortTitle{Pulsar glitch substructure}

\author{\speaker{Richard Dodson}\thanks{Marie Curie Fellow.}\\
        Observatorio Astron\'omico Nacional, Madrid, Espa\~na\\
        E-mail: \email{r.dodson@oan.es}}

\author{Avinash Deshpande\\
        RRI, India\\
        E-mail: \email{desh@rri.res.in}}
\author{Dion Lewis\\
        CSIRO, Australia}
\author{P. McCulloch\\
        University of Tasmania, Australia}

\abstract{Pulsar timing at the Mt Pleasant observatory focused on Vela,
which could be tracked for 18 hours of the day. These nearly continuous
timing records extend over 24 years allowing a great insight into
details of timing noise, micro glitches and other more exotic
effects. It has been found that the spin up for the Vela
pulsar occurs instantaneously to within the uncertainties of the data. The
potential for new, higher resolution data, to unveil insights of the
Neutron Star interiors is discussed. 
}

\FullConference{Bursts, Pulses and 
 Flickering:Wide-field monitoring of the dynamic radio sky\
		 June 12-15 2007\\
		 Kerastari, Tripolis, Greece}

\begin{document}

\section{Introduction}
\label{sec:intro}

Mount Pleasant observatory, just outside Hobart in Tasmania,
Australia, has a 14-metre dish that was dedicated to tracking the
Vela pulsar for two decades. This telescope was able to monitor the
pulsar for eighteen hours every day, and therefore has caught many
glitches `in the act'. As a crosscheck the older, but glitching, PSR
J1644-4559 (B1641-45) was observed for the six hours when Vela was set.  There is no
comparative dataset, and the conclusions we draw puts extremely tight
conditions on the pulsar Equation of State (EOS) placing a number
of constraints on the models. An example of these would be that if the
spin-up is very fast the crust has to have a low moment of inertia,
therefore be very thin, and the coupling between the crust and the
interior super-fluid has to be strong (see, for example, discussion of
these issues in \citet{epst92,bild89}).

Three uncooled receivers were mounted at the prime focus of the
14-metre to allow continuous dispersion measure determination.  Two
used stacked disk, dual polarisation feeds with central frequencies of
635~MHz and 990~MHz additional to a right hand circular helix at
1391~MHz.  Bandwidths were 250~kHz, 800~kHz and 2~MHz respectively,
limiting pulse broadening from interstellar dispersion to less than
1\%.  The output was folded for two minutes giving an integrated pulse
profile of 1344~pulses.  The backend to the 990~MHz receiver also had
incoherent dedispersion across 8 adjacent channels allowing a study of
individual pulses. Results from these systems have been reported,
respectively, in \citet{christ_nat} and \citet{dodson_glitch}.

A new system, based on the PC-EVN VLBI interfaces \citep{dodson_evn},
could produce TOA's with accuracy of the order of 0.1 msec every second
(as opposed to every 10 seconds with the single pulse or 120 seconds
with the multi-frequency systems). This interface was adapted from the
Mets\"ahovi Radio Observatories linux-based DMA, data collector, card
designed for VLBI digital inputs.
The two 40~MHz IFs from the 635~MHz feed provide the two
polarisations, and the data were recorded in a continuous loop two hours
long. This could be halted by the incoherent dedispersion glitch monitoring
program. Coherent dedispersion was to be performed off-line, for the data
segment covering the glitch.

\section{Pulsar Glitches as Transient events}

Transient science is formed of tough targets. By definition the source
is not continuous, and so (in most cases) one does not know when the
event will occur. Even less, in these early days of the field,
one does not really know what many of the targets are. Therefore
neither do we know where to look for them, nor how to interpret the
results. Pulsar glitches, as well as RRAT bursts, and other similar events,
have the advantage that we do know where they occur (on the pulsar
obviously) and they are reasonably repeatable (Vela glitches every
$1000\pm200$ days), and most importantly there exists models of the
Neutron Star EOS which are directly testable with the new systems
being developed. This paper will review the results from the Hobart
system and make suggestions as to which of, and how, the new
instrumentation could be used to provide the clear results that our
systems have hinted at.

\section{The Glitch in 2004}
\label{sec:2004}

\begin{figure}
\centering
  \includegraphics[angle=0,width=0.8\textwidth]{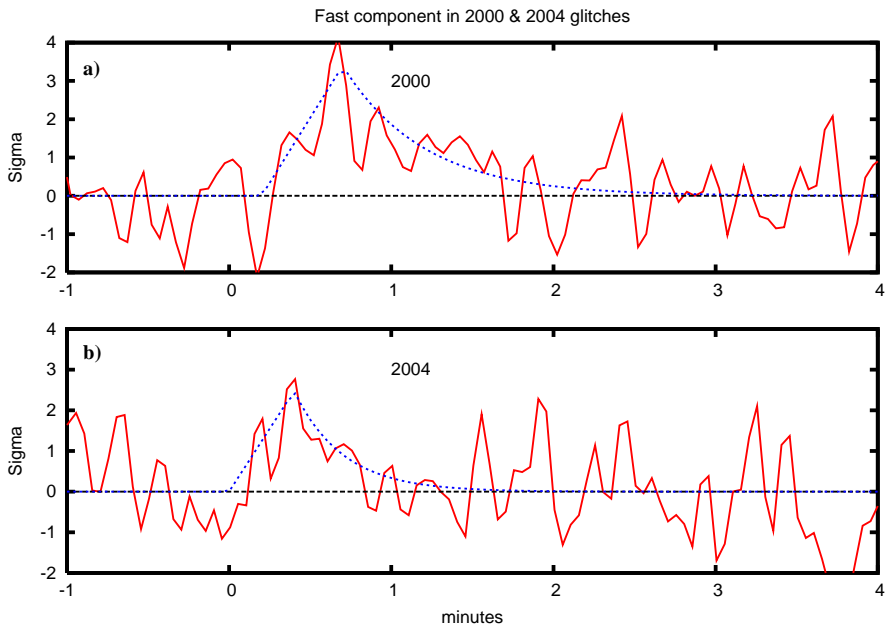}
\caption{The timing residuals plotted as standard deviations, after removing the long scale timing
  models,  against the minutes relative
  to the assumed glitch epoch. The model of a later glitch epoch, and
  a very fast spin down, are overlaid. 1a is for 2000, 1b is for 2004.}
\label{fig:fast}       
\end{figure}

The glitch of 2004 occurred while the telescope was recording
data. Unfortunately the new coherent de-dispersion system was not running
at that moment and the results obtained from the old incoherent system are more or less a repeat of
those in 2000. The instantaneous fractional glitch size was
$2.08\times10^{-6}$. A similar fast decaying term as reported in the
2000 glitch can be seen in the data after the usual model is
subtracted, see Figure \ref{fig:fast}, but it is only a few sigma
above the noise.  This usual model consists of permanent glitch
components in the frequency and frequency deviate, and other
components which are co-temporal jumps in frequency which decay
away. A number of these are required to fit the data and the decay
timescales are denoted with $\tau_n$. Three decaying terms have been
known for sometime and these make up the usual model. In
\citet{dodson_glitch} a fourth short term component was identified.
For a fuller discussion see that paper. In Figure 1 the longer
timescale terms are subtracted (see \citet{dodson_06} for details),
and the residuals are plotted scaled against the RMS. Time zero is the
intercept of the post-glitch model with the pre-glitch model,
i.e. assuming an instant spin-up. The indication of a spin-up would be
negative residuals, of which there is no sign. The positive residuals
seen are modelled as a later glitch epoch and a very fast decaying
term.
Figure 
1a shows the glitch of 2000, where the signal from the fourth component
was clearly above the noise,
Figure 
1b shows the glitch of 2004, where the similar signal is
barely above the noise.
We present it as supporting evidence for the similar signal seen in
the 2000 data, but we are unable to draw more detailed conclusions
from such weak data.

\section{Potential for new observations}

As we have confirmed that at least something similar to the event seen
in 2000 occurred in 2004 we are confident that there is a trail worth
following. Unfortunately it seems very unlikely that the 14-m in
Hobart will continue this work, as it has been upgraded for higher
frequency, non-pulsar, work. Therefore we list what we consider to be
necessary and compare this to facilities available, or soon to be
available. 

\begin{itemize}
\item Base Band Recording
\item Near Real Time Detection
\item Best possible sensitivity
\item Near Continuous Monitoring
\end{itemize}

The first is driven by the ease with which one can collect and record
a very significant portion of the Radio Spectrum, for post
processing. The second is because of the limitations of  storage space that
recording such unprocessed data requires. The next sets the minimum
detection time, and thus is the real limit to the timing resolution of
the data. In the PC-EVN based system we recorded two polarisations of
40\,MHz, at 2 bit resolution which could be stored for several
hours. We used the existing systems to provide the trigger to freeze
the recording, which would have detected a glitch in about 20
minutes. For best possible sensitivity we worked at the lowest
frequency, 635\,MHz. The final is of course because we can not predict
when the event will occur.
The Hobart system, being dedicated, tracked the source for the whole
time it was up: eighteen hours. This is unlikely to be possible, if
this proposed experiment was a piggy back project. However, as the
glitch event we wish to study occurs over seconds to a minute, occasional sampling will not be sufficient. It needs to be tracked
continuously. Phase tracking on several of the proposed SKA
demonstrators does allow the formation of separate beams which could
track the pulsar position, or multiple positions. As the demonstration
of this capability is an important part of many of the proposals this
project could be an attractive one.
Any replacement system will have to be able to recover the data from a
buffer, which could pre-fold the data -- or pulsar gate -- to reduce
the data size. If pre-folding is used the de-dispersion will have to
be applied, and this is quite computationally intensive. If gating is
applied one needs to be very sure of the pulsar phase for an
experiment which will not get much attention until the pulsar
glitches. In our opinion the best option is to collect all the raw
samples. It needs to be possible to access this data, or a subset of
it, to form a monitoring profile. Whilst 40\,MHz was recorded in
Hobart we never processed it in real-time. This could be a major
simplification of the system and it would provide a continuous
health-check. To avoid the computationally expensive de-dispersion a
sub-band would be (digitally) selected, folded and detected for the
pulsar phase. Any sudden (and sustained) deviation from the expected
values would provide the trigger. The size of the sub-band taken,
limited by the dispersion bandwidth, is a function of the sky
frequency used. Selecting the sky frequency requires a compromise of
lowest possible (for best signal to noise) against scatter smearing
which limits the gains from de-dispersion, and the sensitivity in general.

Scattering will be a severe problem for transient science, as very
short signals may well be scattered and spread beyond recovery, see
\cite{jp} in these proceedings. For our case we will not be
additionally hindered by dispersion, as we know the dispersion measure
(DM) for the target. Scattering smears the time signal and the timescale for the scattering,
from \citet{rick_aa}, follows the formula:

{\Large $ \tau_{scatter} =  {z C_G \lambda^4}/{2 \pi^2 c}$}

where $z$ the screen distance, $C_G$ is a constant from the
integration over the path length, $\lambda$ is the wavelength and $c$
the speed of light. In the case of Vela this scales as $9.4\times
(f_{MHz}/300)^{-4}$ msec \citep{kom_72}. If we aimed at a limit of
1\,msec this requires a frequency of 525\,MHz.

We now address these limitations in light of two potential
instruments; MWA and KAT.

MWA (Murchison Widefield Array \citep{ska_path}) is being built in West
Australia and follows the `magic tile' design concept. It is
constructed of active dipoles, and uses phase tracking to follow the
sources on the sky. It has a frequency range of 80 to 300\,MHz, with a
maximum recorded bandwidth of 32\,MHz. Even at the top
frequency the pulsar scattering is 11\% of the profile, which will
reduce the signal to noise and thus the TOA accuracy. The latitude is
$36.6^o$ S. The tracking, however, will be limited as the beam can
only be formed close to the zenith; $\pm25^o$ at 150\,MHz.  The
T$_{\rm ant}$ is not a specification which is easy to derive, but will
be around the sky temperature, which we approximate as 150\,K at
300\,MHz. The effective area of each tile will be $\sim 10{\rm m}^2$,
and there will eventually be 500 tiles. We have assumed 16 tiles are
combined to give a SEFD of 4000\,Jy , two polarisations of 32\,MHz and
using 7\,Jy for the flux we can expect a SNR of 42 in one second.

KAT (Karoo Array Telescope \citep{ska_path}) is being built in South Africa
and follows the small D-large N design concept. It will be constructed
of $\sim80$ dishes of 12m. KAT-7, a rollout prototype, is being built
now. The frequency range is 700 (or perhaps 500) to 1700 (or perhaps
2500) MHz. There are no available specifications for the sensitivity
or the bandwidth. Nevertheless we can assume some typical values for
these parameters. 700\,MHz will be the best for the project we have in
mind, with the scattering limited to 0.3 msec for Vela. Assuming a
SEFD of 2000\,Jy (or a T$_{\rm ant}$ of about 70\,K), two
polarisations of 32\,MHz and using 2.5\,Jy for the flux we can expect
a SNR of 65 in one second. The latitude, $25.9^o$ S, is well placed
for tracking Vela (and other sources towards the galactic centre).  The
disadvantage is that we would require a dedicated dish -- the single
beam option is not possible. This may preclude the project, unless
there was considerable downtime in the testing phase of KAT-7.

\section{Conclusions}
\label{sec:con}

The Vela pulsar was timed for more than twenty years, and provided new
insights into the pulsar EOS by, for example, providing very low
limits for the spin-up time and therefore the crust thickness.  There
is clearly a need to observe a glitch with higher sensitivity and time
resolution to investigate both the fast decay term and to detect the
spin-up. Both of these values relate directly to the pulsar EOS and
will provide rich fodder for theoretical analysis by allowing the
measurement of the moment of inertia of the crust. With the
termination of the observation program this effect is a ripe target
for the next generation of pulsar telescopes that could monitor a
large number of targets simultaneously with new beam forming
techniques. Two next generation telescopes were considered, MWA and
KAT. Both would be capable of detecting the pulsar with sufficient
signal to noise to track the very short term behaviour of the pulsar
after a glitch. The MWA, however, would have trouble tracking the pulsar for
a large fraction of the day, and therefore we feel the most suitable
instrument for continuing this work will be subsections of KAT-7, when
not being used for SKA development.

%
\begin{figure}
\centering
  \includegraphics[angle=0,width=0.8\textwidth]{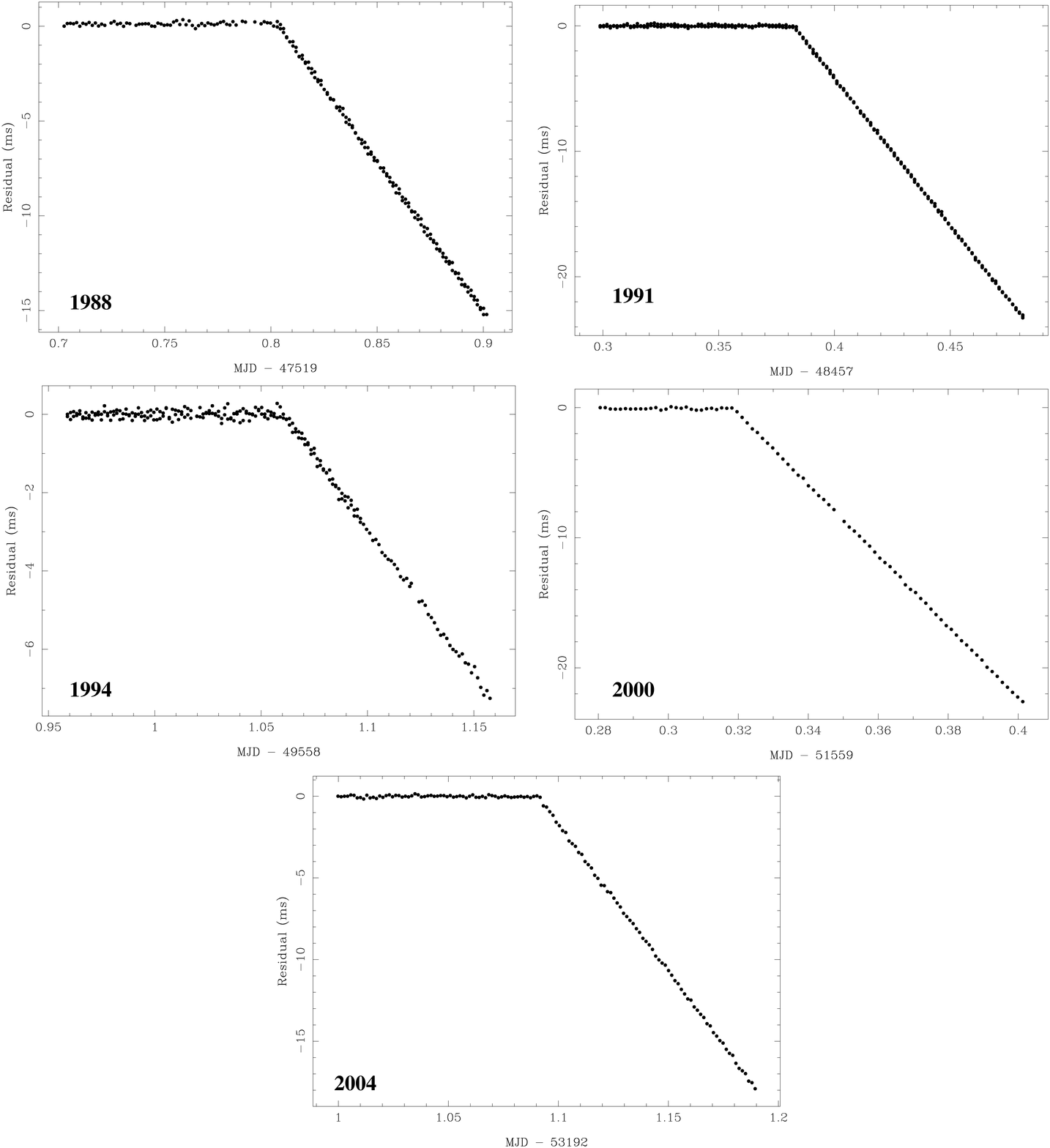}
\caption{Plots of all the glitches of Vela directly observed. The
  residuals are in milliseconds are plotted against time in days.}
\label{fig:glitches}       
\end{figure}


\begin{thebibliography}{99}
%
\bibitem{bild89}
{{Bildsten}, L. and {Epstein}, R.~I.}, {\em Superfluid dissipation time
scales in neutron star crusts} 
\newblock Astrophys. J. \textbf{342}, 951--957 (1989)

%
\bibitem{dodson_evn}
{Dodson}, R., {Tingay}, S., {West}, C., {Phillips}, C., {Tzioumis}, A.K.,
  {Ritakari}, J., {Briggs}, F., {\em The Australian experience with the PC-EVN
  recorder}.
\newblock In: R.~{Bachiller}, F.~{Colomer}, J.F. {Desmurs}, P.~{de Vicente}
  (eds.) EVN on New Developments in VLBI, pp. 253--255 (2004) 

%
\bibitem{dodson_glitch}
{Dodson}, R., {McCulloch}, P.M., {Lewis}, D.R. {\em High Time Resolution
  Observations of the January 2000 Glitch in the Vela Pulsar}.
\newblock Astrophys. J. \textbf{564}, L85--L88 (2002)

%
\bibitem{dodson_06}
{Dodson}, R., {Lewis}, D.R., {McCulloch}, P.M., {\em Two decades of pulsar timing of Vela}.
\newblock ApSS \textbf{308}, 585 (2007)

%
\bibitem{epst92}
{{Epstein}, R.~I. and {Baym}, G.}, {\em Vortex drag and the spin-up time
scale for pulsar glitches} 
\newblock Astrophys. J. \textbf{387}, 276--287 (1992)






%
\bibitem{ska_path} Greenwood, C., 
{\em Status of Pathfinder Telescopes and Design Studies}
\newblock Int. SKA Project Office, \textbf{1.7.1}, 2007

%
\bibitem{kom_72} Komesaroff, M.~M., 
Hamilton, P.~A., Ables, J.~G., {\em Linear polarization and spectrum of 
PSR 0833-45 and the effects of scattering.}
\newblock Australian Journal of Physics 
\textbf{25}, 759, 1972

%
\bibitem{jp} Macquart, J.P., 
{\em Limits on the Detection of Transients Imposed by Scattering}
\newblock  PoS(Dynamic2007) \textbf{022}, 2008, [{\tt arXiv:0711.2535}]

%
\bibitem{christ_nat}
{McCulloch}, P.M., {Hamilton}, P.A., {McConnell}, D., {King}, E.A., {\em The VELA
  glitch of Christmas 1988}.
\newblock Nature \textbf{346}, 822--824 (1990)

%
\bibitem{rick_aa} Rickett, B.~J., 
{\em Interstellar scattering and scintillation of radio waves}
\newblock ARAA, \textbf{15}, 479, (1977)




\end{thebibliography}
\end{document}